\documentclass[12pt,prb,amsfonts,amssymb,amsmath,aps,titlepage]{revtex4-1}
\usepackage{graphicx}
\begin{document}
\baselineskip 24pt

\def\ket#1{|#1\rangle}
\def\bra#1{\langle#1|}
\def\av#1{\langle#1\rangle}

\title{Oscillator tunneling dynamics in the Rabi model}

\author{E. K. Irish}
\affiliation{SUPA, School of Physics and Astronomy, University of St. Andrews, St. Andrews, KY16 9SS, UK}
\email{eki2@st-andrews.ac.uk}
\author{J. Gea-Banacloche}
\affiliation{Department of Physics, University of Arkansas, Fayetteville, AR 72701, USA}
\email{jgeabana@uark.edu}
\begin{abstract}
The familiar Rabi model, comprising a two-level system coupled to a quantum harmonic oscillator, continues to produce rich and surprising physics when the coupling strength becomes comparable to the individual subsystem frequencies. We construct approximate solutions for the regime in which the oscillator frequency is small compared to that of the two-level system and the coupling strength matches or exceeds the oscillator frequency. Relating our fully quantum calculation to a previous semi-classical approximation, we find that the dynamics of the oscillator can be considered to a good approximation as that of a particle tunneling in a classical double-well potential, despite the fundamentally entangled nature of the joint system. We assess the prospects for observation of oscillator tunneling in the context of nano- or micro-mechanical experiments and find that it should be possible if suitably high coupling strengths can be engineered.
\end{abstract}

\maketitle

\section{Introduction}

When Jaynes and Cummings introduced their theory of the molecular beam maser in 1963,\cite{Jaynes1963} they presumably had no idea that their modest quantum model comprising a two-level system coupled to a quantized harmonic oscillator would still be the subject of active research fifty years later. The sheer simplicity of the model, and the fact that it features the interaction of two of the most basic quantum systems, have allowed it to be applied to numerous experimental systems beyond the original maser setting. For many years the primary experimental realization was in cavity quantum electrodynamics (cavity QED), in which an atom interacts with the electromagnetic field inside an optical or microwave cavity.\cite{Haroche1989,Miller2005} In the absence of a full analytical solution, theoretical treatments were dominated by the rotating-wave approximation (RWA) applied by Jaynes and Cummings, which provides an excellent description of the energies and eigenstates within the parameter regimes accessible in cavity QED experiments.\cite{Shore1993} However, recent years have seen a proliferation of engineered quantum systems whose behavior is well described by the same model, including several types of superconducting qubits coupled to microwave waveguide resonators,\cite{Fink2008,Niemczyk2010,Pirkkalainen2013} LC resonators,~\cite{Forn-Diaz2010} or mechanical resonators~\cite{LaHaye2009, Suh2010,O'Connell2010,Pirkkalainen2013}; intersubband transitions in semiconductor microcavities~\cite{Anappara2009}; photochromic molecules in metallic cavities~\cite{Schwartz2011}; quantum wells coupled to split-ring resonators~\cite{Scalari2012}; and a photonic analogue in waveguide superlattices~\cite{Crespi2012}. Such systems are capable of accessing very different parameter regimes, in particular much larger coupling strengths than those available in cavity QED, which has in turn inspired a revival of theoretical interest in the model.

The Hamiltonian for the two-level system \textemdash oscillator system may be written as
\begin{equation}
H=\frac{\hbar\Omega}{2}\,\sigma_x +\hbar\omega_0 a^\dagger a + \hbar\lambda\sigma_z(a^\dagger + a) ,
\label{e1}
\end{equation}
where $\Omega$ is the energy difference between the levels of the two-level system (for which we will use the term `qubit'), $\omega_0$ is the frequency of the oscillator, and $\lambda$ is the strength of the coupling between them. This is often called the (quantum) Rabi model or the single-mode spin-boson Hamiltonian, since the term ``Jaynes-Cummings model'' has become synonymous with the RWA.  

In recent years considerable progress has been made in understanding the full model without the RWA. Formally exact mathematical solutions have finally been found~\cite{Braak2011,Zhong2013}, after many years of uncertainty as to whether such solutions even existed. At a more intuitive level, the model may be divided into several regimes depending on the ratios $\Omega/\omega_0$ and $\lambda/\omega_0$. It is now well understood that the RWA is suitable for near-resonance, $\Omega/\omega_0 \approx 1$, and small coupling, $\lambda/\omega_0 \lesssim 0.1$. Away from resonance, the small coupling limit $\lambda/\omega_0 \ll 1$ may be treated with standard perturbation theory. Hence current theoretical research is mostly focused on the very strong coupling limit, $\lambda/\omega_0 \gtrsim 0.1$.

For the regime in which $\Omega/\omega_0 < 1$, an excellent approximation may be obtained from lowest-order degenerate perturbation theory in the basis of states obtained by setting $\Omega = 0$ in Eq.~\eqref{e1}, for which we will use the term ``adiabatic approximation''~\cite{Irish2005}. The same expressions can also be derived by other methods \cite{Schweber1967,Graham1984, Crisp1992,Sandu2009,Ashhab2010}. This approximation provides good agreement with the numerically determined eigenstates and energies for arbitrary coupling values when $\Omega/\omega_0 \ll 1$. Although the adiabatic approximation is derived on the assumption of small $\Omega$, qualitative agreement with numerics persists over the full range of coupling strengths as long as $\Omega/\omega_0 \lesssim 1$ \cite{Ashhab2010}. A number of techniques have been proposed to make quantitative corrections to the states and energies but the corrections generally prove to be small, reaching their largest values for the lowest levels and as $\Omega$ approaches $\omega_0$ \cite{Liu2009,Casanova2010, Hausinger2010}.

The situation becomes more complicated for $\Omega/\omega_0 \geq 1$. Variational treatments based on the form of the adiabatic approximation have been shown to improve the ground-state energies and wavefunctions for larger values of $\Omega$~\cite{Shore1973,Stolze1990,Hwang2010,Bera2013}. However, the excited states cannot generally be treated in this way. As $\Omega$ becomes equal to or greater than $\omega_0$, resonances occur in the excited-state spectrum at small coupling values~\cite{Feranchuk1996, Amniat-Talab2005,Irish2007}; the effect of this is that unphysical level crossings appear in the adiabatic energy levels. This represents a qualitative change in the eigenstates and energies which must be treated by different methods~\cite{Amniat-Talab2005,Pereverzev2006,Irish2007,Feranchuk2009,Liu2009,Hausinger2010,Ashhab2010,Hwang2010}. At larger coupling, the effect of these resonances is no longer observed and the adiabatic approximation still works well, both qualitatively and quantitatively~\cite{Ashhab2010, Wolf2013}. This regime has been termed the `deep strong coupling' limit in the literature~\cite{Casanova2010}, and the criterion usually given for achieving this limit is $\lambda/\omega_0 \sim 1$ \cite{Casanova2010, Wolf2013}.

It has been known for quite some time that a critical point at $\lambda_c = \tfrac{1}{2} \sqrt{\Omega \omega_0}$ exists in some semiclassical approximations in the slow-oscillator limit, at which the nature of the system’s wavefunction undergoes a sharp change~\cite{Graham1984,Ashhab2010,Hwang2010}. The behavior of various system observables in this region has been studied in detail by Ashhab~\cite{Ashhab2013}. The existence of the critical point implies that the criterion for achieving the deep strong coupling limit, when the adiabatic approximation becomes a good description, must be modified. This leaves a wide intermediate coupling region, including the critical point itself, in which the behavior of the system may be expected to deviate strongly from both weak-coupling theory and the adiabatic approximation. There is evidence that squeezing of the oscillator state is involved to some extent near the critical point. A number of authors have incorporated squeezing into variational studies of the ground state~\cite{Shore1973,Chen1989,Chen1989b,Stolze1990,Hwang2010}, and numerical calculations have shown how the squeezing parameter in the ground state scales with $\Omega/\omega_0$ and $\lambda/\omega_0$ \cite{Ashhab2010,Ashhab2013}. The dynamics of squeezing in wavepacket evolution has been discussed by Sandu et al. ~\cite{Sandu2003} and Larson~\cite{Larson2007}. However, the properties of the system, particularly its dynamics, in this intermediate coupling regime are still far from fully understood. 

In this paper we address the case of $\Omega > \omega_0$, for couplings above the critical point but below the region in which the adiabatic approximation becomes valid. We present an approximate solution for both ground and excited states and show that, under certain conditions, it predicts tunneling-like behavior for the oscillator. In Section 2 we derive the approximation for the energies and wavefunctions of the system, discuss its range of validity, and calculate dynamics of both oscillator and qubit observables. Section 3 contains a discussion of several ways to interpret the dynamical behavior in terms of effective double-well potentials. Prospects for experimental observations of the tunneling dynamics in nanomechanical systems are outlined in Section 4, and we draw some brief conclusions in Section 5.

\section{Approximate solution in the slow-oscillator limit}

The Hamiltonian \eqref{e1} can also be expressed as
\begin{equation}
H = \frac{\hbar\Omega}{2}\bigl(\ket{+z}\bra{-z} + \ket{-z}\bra{+z}\bigr) + H_L \ket{+z}\bra{+z} + H_R \ket{-z}\bra{-z} -\frac{\hbar \lambda^2}{\omega_0}
\label{e2}
\end{equation}
where $H_L$ and $H_R$ are the Hamiltonians of oscillators displaced to the left and right, respectively:
\begin{equation}
H_{L,R} =\hbar\omega_0\left(a^\dagger \pm \lambda/\omega_0\right)\left(a \pm \lambda/\omega_0\right) .
\label{e3}
\end{equation}
If the qubit energy term (first term in (\ref{e1}) and (\ref{e2})) is neglected, the energy eigenstates are simply displaced number states of the oscillator, associated with either the $\ket{+z}$ state of the qubit (left-displaced) or the $\ket{-z}$ state (right-displaced). This is the starting point of the adiabatic approximation~\cite{Irish2005}, in which the $\Omega$ term is treated perturbatively.  Here, however, we are interested in studying the case $\Omega/\omega_0 > 1$.

One approach that has been taken previously is to treat the displacement of the oscillators as a variational parameter and minimize the ground state energy with respect to the displacement. While this does improve upon the ground state of the adiabatic approximation, it is not particularly accurate in the intermediate coupling regime and, furthermore, produces unphysical `kinks' in the values of various observables including the energy~\cite{Shore1973,Bera2013}. 

We begin instead with a variational test function in which both the oscillator displacement and the rotation of the qubit state are taken as variational parameters:
\begin{equation}
\ket{\psi_0} = \ket\alpha \otimes\left(\cos\frac\theta 2 \ket{+x} + \sin\frac\theta 2 \ket{-x} \right)
\label{e4}
\end{equation}
where $\ket{\alpha}$ is a coherent state, with wavefunction, in the coordinate representation,
\begin{equation}
\av{q | \alpha} = \left(\frac{m\omega_0}{\pi\hbar}\right)^{1/4}\exp\left(-\frac{m\omega_0}{2\hbar}(q - q_{\alpha})^2\right) = \left(\frac{m\omega_0}{\pi\hbar}\right)^{1/4}\exp\left[-\left(\sqrt{\frac{m\omega_0}{2\hbar}}\,q - \alpha\right)^2\right] ,
\label{e5}
\end{equation}
where $q_{\alpha} \equiv \alpha\sqrt{2\hbar/m\omega_0}$ with $m$ the mass of the oscillator. Assuming $\alpha$ to be real, the expectation value of $H$ in the state $\ket{\psi_0}$ is
\begin{equation}
\bra{\psi_0}H\ket{\psi_0} = \frac{\hbar\Omega}{2}\cos\theta + 2\hbar\lambda\alpha \sin\theta +\hbar\omega_0 \alpha^2 .
\label{e6}
\end{equation}
We may search for the ground state energy by minimizing \eqref{e6} with respect to the parameters $\alpha$ and $\theta$.  If $4\lambda^2/\Omega\omega_0 <1$ there is only one minimum, at $\alpha=0$ and $\theta = \pi$, but if instead the condition
\begin{equation}
\frac 1\epsilon \equiv \frac{4\lambda^2}{\Omega\omega_0} >1
\label{e8}
\end{equation}
holds, there are two minima, given by $\alpha = \pm \alpha_0$ and $\theta = \pm\theta_0$, with
\begin{equation}
\cos\theta_0 = -\frac{\Omega\omega_0}{4\lambda^2} = -\epsilon .
\label{e9} 
\end{equation}
and
\begin{equation}
\alpha_0 = -\frac{\lambda}{\omega_0}\sin\theta_0 = -\frac{\lambda}{\omega_0}\sqrt{1-\left(\frac{\Omega\omega_0}{4\lambda^2}\right)^2} = -\frac{\lambda}{\omega_0}\sqrt{1-\epsilon^2}
\label{e10}
\end{equation}
In the remainder of this paper, we will assume that condition \eqref{e8} always holds.  Also, for definiteness, assume that $\theta_0$ is chosen so that $\sin\theta_0 \ge 0$, which makes $\alpha_0 \le 0$.  Then there are two degenerate states, $\ket{\psi_{0,L}} = \ket{\alpha_0}(\cos\frac{\theta_0}{2}\ket{+x} + \sin\frac{\theta_0}{2}\ket{-x})$ (left-displaced) and $\ket{\psi_{0,R}} = \ket{-\alpha_0}(\cos\frac{\theta_0}{2}\ket{+x} - \sin\frac{\theta_0}{2}\ket{-x})$ (right-displaced), with energy
\begin{equation}
E(\alpha_0,\theta_0) = -\frac{\hbar\lambda^2}{\omega_0} -\frac{\hbar\Omega^2\omega_0}{16 \lambda^2}  = -\frac{\hbar\lambda^2}{\omega_0} \left(1+\epsilon^2\right) .
\label{e11}
\end{equation}
Neither of these functions, however, are eigenstates of the parity operator $\hat{P} = \exp[i\pi(a^\dagger a + \frac 1 2 \sigma_x - \frac{1}{2})]$ that commutes with $H$; see, e.g., Ref.~\onlinecite{Graham1984}. States of definite parity may be obtained by taking the superpositions
\begin{equation}
\ket{\Phi_{\pm,0}} = \frac{1}{\sqrt 2 {\cal N}_{\pm,0}} \bigl(\ket{\psi_{0,L}} \pm \ket{\psi_{0,R}} \bigr)
\label{e12}
\end{equation}
where ${\cal N}_{\pm,0}$ is an appropriate normalization factor, which is needed because $\ket{\psi_{0,L}}$ and $\ket{\psi_{0,R}}$ are not, in general, orthogonal.  Indeed, using ${\cal N}_{\pm,0}=\sqrt{1\mp \epsilon \, e^{-2\alpha_0^2}}$ (see Eqs.~(\ref{e15}) and (\ref{e17}) below), one can show that $\bra{\Phi_{-,0}}H\ket{\Phi_{-,0}} < E(\alpha_0,\theta_0)$ as long as (\ref{e8}) holds, and hence $\ket{\Phi_{-,0}}$ is a better approximation to the ground state of the system.  

This suggests that we consider the set of states defined by
\begin{equation}
\ket{\Phi_{\pm,N}} = \frac{1}{\sqrt 2 {\cal N}_{\pm,N}} \bigl(\ket{\psi_{N,L}} \pm \ket{\psi_{N,R}} \bigr)
\label{e13}
\end{equation}
in terms of the left- and right-displaced states
\begin{align}
\ket{\psi_{N,L}} &= \hat D[\alpha_0]\ket N \otimes \left(\cos\frac{\theta_0}{2} \ket{+x} + \sin\frac{\theta_0}{2} \ket{-x} \right) \cr
\ket{\psi_{N,R}} &= \hat D[-\alpha_0]\ket N \otimes \left(\cos\frac{\theta_0}{2} \ket{+x} - \sin\frac{\theta_0}{2} \ket{-x} \right)
\label{e14}
\end{align}
and the normalization coefficients
\begin{equation}
{\cal N}_{\pm,N} = \left[1\mp \epsilon \, e^{-2(\lambda^2/\omega_0^2)(1-\epsilon^2)} L_N\left(\frac{4\lambda^2}{\omega_0^2}(1-\epsilon^2)\right)\right]^{1/2} .
\label{e15}
\end{equation}
In Eqs.~(\ref{e14}), $\hat D[\alpha] \equiv \exp[\alpha(a^\dagger - a)]$ is a displacement operator, and the $\hat D[\pm\alpha_0]\ket N$ are displaced number states, which (for given $\alpha_0$) constitute two equivalent, alternate bases for the oscillator Hilbert space.  The generic displaced number state $\hat D[\alpha]\ket N$ can also be written as $\ket{\alpha,N}$, and the inner product of number states displaced in different directions is given by a Laguerre polynomial:
\begin{equation}
\av{-\alpha,N|\alpha,N} = e^{-2\alpha^2} L_N(4\alpha^2) .
\label{e16}
\end{equation}
For large $N$, the right-hand side of Eq.~(\ref{e16}) decays as $1/(N^{1/4}\alpha^{1/2})$.  

The states $\ket{\psi_{N,L}}$ and $\ket{\psi_{N,R}}$ may be thought of as arising from the same variational calculation that yielded $\ket{\psi_{0,L}}$ and $\ket{\psi_{0,R}}$, only starting from a field state which is a displaced number state (note that a coherent state is equivalent to a displaced vacuum state).  The calculation yields the same optimal values $\alpha_0$ and $\theta_0$ for all values of $N$.  All the $\ket{\psi_{N,L}}$ are orthogonal for different $N$, as are the $\ket{\psi_{N,R}}$, but the left- and right-displaced states, as indicated above, are not orthogonal to each other. For the same $N$, $\ket{\psi_{N,L}}$ and $\ket{\psi_{N,R}}$ are degenerate in energy, and this, together with the parity considerations, is what leads us to consider the positive and negative superpositions given by Eq.~\eqref{e13}.  

It is easy to see that the states $\ket{\Phi_{\pm,N}}$, taken together, form a complete but not an orthogonal set.  The parity of the state $\ket{\Phi_{\pm,N}}$ is $\pm(-1)^N$, and states of opposite parity are orthogonal, but states of the same parity in general are not; their overlap is proportional to terms of the form $\av{-\alpha,N|\alpha,M}$ which are given by associated Laguerre polynomials. However, just like (\ref{e16}), the overlap decreases at least as fast as $1/\alpha^{1/2}$ for large $N$ and $M$.  Recalling expression (\ref{e10}) for the displacement $\alpha_0$, it appears that these states can provide an approximately orthogonal basis in the limit $\lambda \gg \omega_0$. Additionally, the overlap terms are all suppressed by a factor $\cos\theta = -\epsilon$ (see Eq.~(\ref{e9})), which is also small for large coupling constant $\lambda$.

\begin{figure}[htbp]
\begin{center}
\includegraphics[]{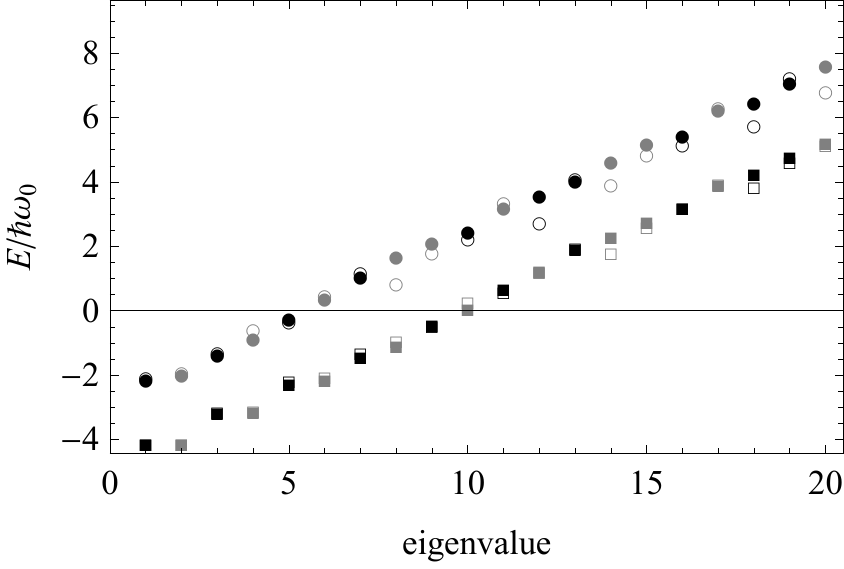}
\caption{Energy values for $\Omega = 3\omega_0$ given by the approximation of Eq.~\eqref{e17} (open symbols) and by numerical diagonalization of the full Hamiltonian (solid symbols). Squares (circles) correspond to $\lambda = 2\omega_0$ ($\lambda = 1.3\omega_0$) and black (grey) denotes negative (positive) parity.}
\label{energies}
\end{center}
\end{figure}

In this limit, then, we shall approximate the energy eigenvalues of $H$ by the expectation values $E_{\pm,N} = \bra{\Phi_{\pm,N}}H\ket{\Phi_{\pm,N}}$, that is to say, the diagonal elements of the Hamiltonian in the approximately orthogonal basis (\ref{e13}).  The justification is, again, that the off-diagonal elements can be made small for sufficiently large $\alpha_0$.  Explicitly, the approximate energies $E_{\pm,N}$ are given by
\begin{equation}
\begin{split}
E_{\pm,N} &= \frac{\hbar}{1 \mp \epsilon e^{-2\alpha_0^2}L_N(4\alpha_0^2)} \bigg[-\frac{\Omega}{2} \epsilon + N\omega_0 - \frac{\lambda^2}{\omega_0}\left(1 - \epsilon^2\right) \\
& \qquad \pm \left(\frac{\Omega}{2} - N\omega_0\epsilon + \frac{\lambda^2}{\omega_0} \epsilon(1 - \epsilon^2) \right)e^{-2\alpha_0^2}L_N(4\alpha_0^2) \bigg]  \\
&= \frac{\hbar}{1 \mp \epsilon e^{-2\alpha_0^2}L_N(4\alpha_0^2)} \bigg[-\frac{\Omega}{2} \left( \epsilon \mp e^{-2\alpha_0^2}L_N(4\alpha_0^2) \right) \bigg] \\
&\quad + \hbar N\omega_0 - \hbar \frac{\lambda^2}{\omega_0}\left(1 - \epsilon^2\right) .
\end{split}
\label{e17}
\end{equation}
To give an idea of how well this approximation works, Fig.~\ref{energies} shows the first twenty eigenvalues of $H$ calculated by numerical diagonalization of the full Hamiltonian together with the corresponding values from Eq.~(\ref{e17}), for $\Omega = 3\omega_0$ and two coupling values, $\lambda = 2\omega_0$ and $\lambda = 1.3\omega_0$.  The first case, where agreement is overall quite good, corresponds to $\epsilon = 0.19$ and $\alpha_0 = -1.96$, whereas the second case corresponds to $\epsilon = 0.44$ and $\alpha_0 = -1.16$.  Even though the latter case pushes the approximation to its limits, we note that the agreement is still fairly good for the lowest eigenvalues: the approximate values for the lowest doublet are $(-2.10,-1.94)$ whereas the exact ones are $(-2.17,-2.01)$, a difference of less than $4$ percent.


\begin{figure}[htbp]
\begin{center}
\includegraphics[]{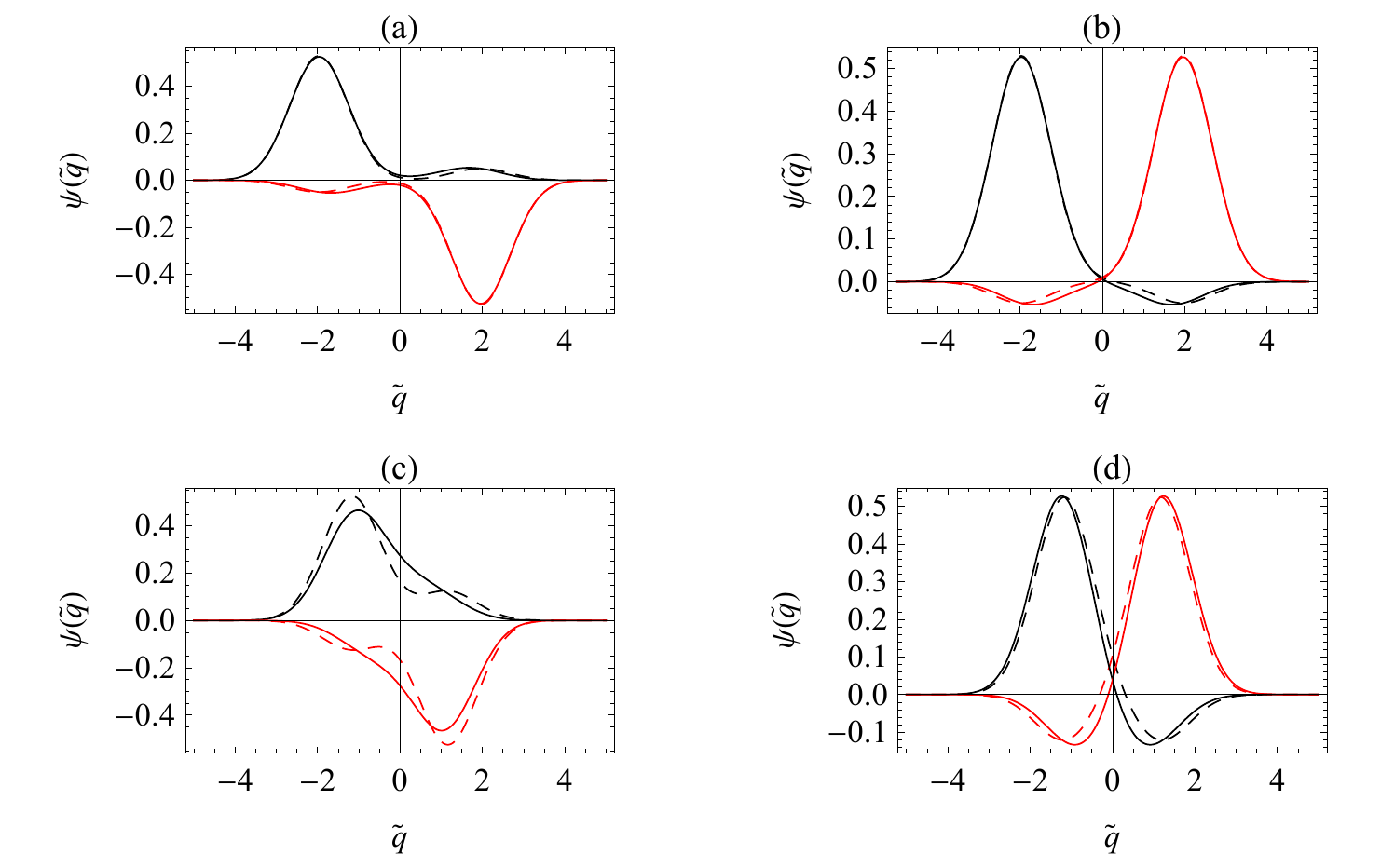}
\caption{Oscillator wavefunctions projected along $\ket{+z}$ (black) and $\ket{-z}$ (red) in position space with $\Omega = 3\omega_0$. Position is plotted as a function of $\tilde{q} = \sqrt{2 \hbar/m \omega_0} q$. Solid lines are obtained by numerical diagonalization of the full Hamiltonian, while dashed lines correspond to Eq.~\eqref{e12}. Shown are the ground state (a) and first excited state (b) for $\lambda = 2\omega_0$ and the ground state (c) and first excited state (d) for $\lambda = 1.3\omega_0$.}
\label{wavefuncs}
\end{center}
\end{figure}

The approximation is also reasonably good for the eigenfunctions. Figure~\ref{wavefuncs} shows the projections of the ground and first excited state wavefunctions along $\ket{\pm z}$, plotted in position space, for $\lambda = 2\omega_0, 1.3\omega_0$. These plots demonstrate clearly the need to treat the spin rotation as a variational parameter. The right (left) peak in the $\ket{+z}$ ($\ket{-z}$) projection would be absent for $\theta_0$ fixed at $\pi/2$, as it would be in a standard variational displacement calculation. A similar, albeit more general, ansatz for the ground-state energy of the spin-boson model has been studied by Bera \textit{et al.}~\cite{Bera2013}, who have shown that including `antipolaronic' terms, in which the oscillator is displaced in the opposite direction to that predicted by the adiabatic approximation, provides a significant improvement over the simple variational displacement. Although it is clear from Fig.~\ref{wavefuncs}(c,d) that the approximation is starting to break down for $\lambda = 1.3\omega_0$, the overlap in amplitude  of $\ket{\Phi_{-,0}}$ with the numerically calculated ground state is still $0.977$, and the overlap of $\ket{\Phi_{+,0}}$ with the next-highest numerically calculated state is $0.991$. 

More importantly, the approximation we have developed here offers an intuitive means of understanding the dynamics of the system. As shown in Eq.~(\ref{e12}), both $\ket{\Phi_{-,0}}$ and $\ket{\Phi_{+,0}}$ involve a superposition of displaced coherent states, $\ket{\pm\alpha_0}$.  Ignoring the small difference between ${\cal N}_{\pm,0}$ and 1, one expects $\ket{\Phi_{-,0}} + \ket{\Phi_{+,0}}$ to be proportional to the left-displaced coherent state $\ket{\alpha_0}$ (recall $\alpha_0$ is negative), and $\ket{\Phi_{-,0}} - \ket{\Phi_{+,0}}$ to be proportional to the right-displaced coherent state $\ket{-\alpha_0}$. Suppose that the system is initially prepared in the state
\begin{equation}
\ket{\Psi(t=0)} = \ket{\psi_{0,L}} = \ket{\alpha_0}\left(\cos\frac{\theta_0}{2}\ket{+x} +\sin\frac{\theta_0}{2}\ket{-x} \right) \simeq \frac{1}{\sqrt 2} \bigl(\ket{\Phi_{-,0}} + \ket{\Phi_{+,0}}\bigr) .
\label{e18}
\end{equation}
The time evolution can be approximated by
\begin{equation}
\ket{\Psi(t)} = \frac{e^{-iE_{-,0}t/\hbar}}{\sqrt 2} \bigl(\ket{\Phi_{-,0}} + e^{-i\Delta\omega t}\ket{\Phi_{+,0}}\bigr)
\label{e19}
\end{equation}
where $\Delta\omega$ is the frequency difference for the ground state doublet:
\begin{equation}
\Delta\omega = \frac{E_{+,0}-E_{-,0}}{\hbar} = \frac{\Omega (1-\epsilon^2) e^{-2 \alpha_0^2}}{1-\epsilon^2 e^{-4 \alpha_0^2}} \simeq \Omega (1-\epsilon^2) e^{-2 \alpha_0^2} ,
\label{e20} 
\end{equation}
where in the last expression we have neglected higher powers of $e^{-2\alpha_0^2}$.  As indicated above, at the initial time $t=0$ the state (\ref{e19}) corresponds to the oscillator being localized mostly on the left (coherent state $\ket{\alpha_0}$), whereas at the time $t=\pi/\Delta\omega$ it will be localized on the right (coherent state $\ket{-\alpha_0}$). At the intermediate time $t=\pi/2\Delta\omega$ it will have a doubly peaked position probability distribution corresponding to a superposition of two coherent states.  This is standard tunneling motion, as in the classic double-well potential. 

\begin{figure}[htbp]
\includegraphics[]{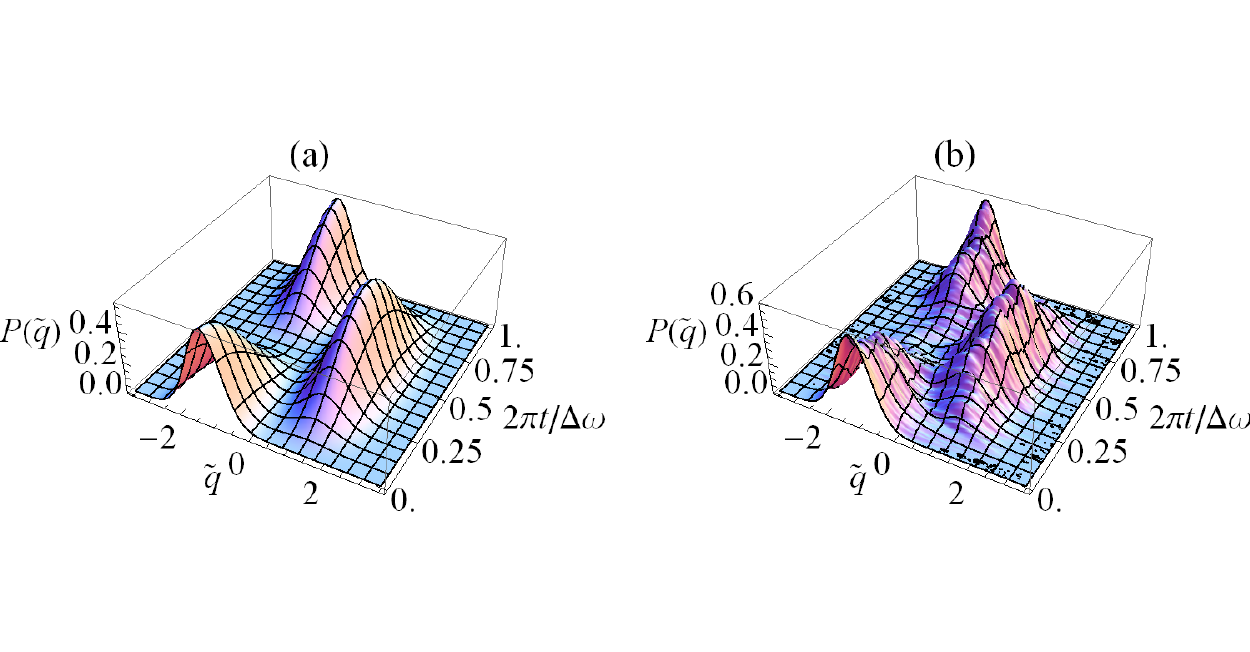}
\caption{Time evolution of the oscillator probability distribution in position space with $\Omega = 3 \omega_0$ and $\lambda = 1.3 \omega_0$, with the initial condition $\Psi(t=0) = \psi_{0,L}$. Position is plotted as a function of $\tilde{q} = \sqrt{2 \hbar/m \omega_0} q$; the time axis is scaled by the tunneling period $2 \pi/\Delta \omega$ where $\Delta \omega$ is given by Eq.~\eqref{e20}. (a) Approximate evolution given by Eq.~\eqref{e19}; (b) numerically calculated evolution with the full Hamiltonian.}
\label{oscdynamics}
\end{figure}

Figure~\ref{oscdynamics} compares the dynamics of the oscillator's position-space probability distribution given by (a) the approximation of Eq.~\eqref{e20} and (b) a numerical calculation with the full Hamiltonian, for $\Omega = 3\omega_0$ and $\lambda = 1.3 \omega_0$. Even for these parameters which push its limits of validity, the approximation captures the coarse-grained dynamics of the oscillator surprisingly well. The characteristic tunneling behavior, in which the probability to be localized on the left or right oscillates in time while the probability to be found at the origin remains negligible, can clearly be seen in both the approximation and the full numerical calculation. Note that the approximate result (\ref{e20}) for the frequency of the tunneling motion (here $0.164$, in units of $\omega_0$) agrees well with the numerically calculated one (here $0.156$).

\begin{figure}[htbp]
\includegraphics[]{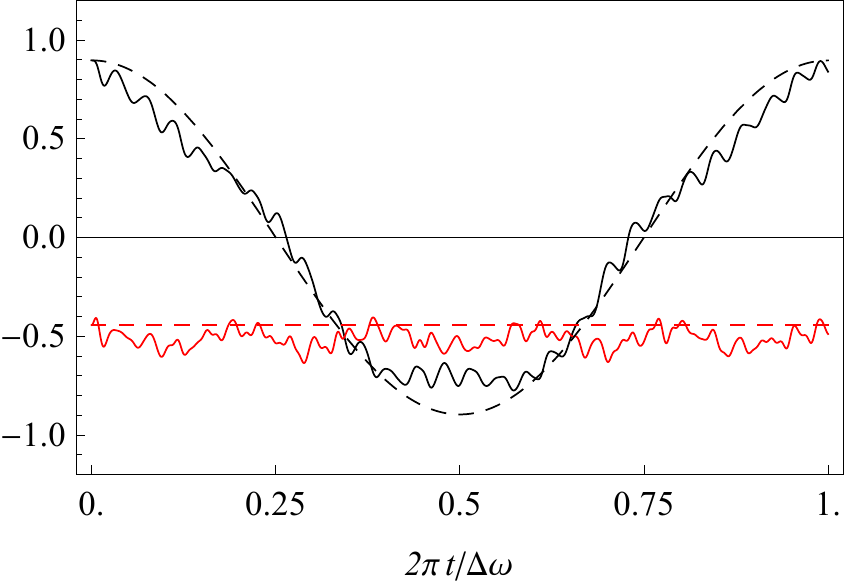}
\caption{Time evolution of $\av{\sigma_z}$ (black) and $\av{\sigma_x}$ (red) with $\Omega = 3 \omega_0$ and $\lambda = 1.3 \omega_0$, with the initial condition $\Psi(t=0) = \psi_{0,L}$. The time axis is scaled by the tunneling period $2 \pi/\Delta \omega$ where $\Delta \omega$ is given by Eq.~\eqref{e20}. Dashed lines: approximate evolution given by Eq.~\eqref{e19}; solid lines: numerically calculated evolution with the full Hamiltonian.}
\label{spindynamics}
\end{figure}

The corresponding dynamics of the qubit observables $\av{\sigma_z}$ and $\av{\sigma_x}$ are plotted in Fig.~\ref{spindynamics}. Expressions for these quantities are easily calculated from Eq.~\eqref{e19}, giving $\av{\sigma_z} = \sin \theta_0 \cos(\Delta \omega t)$ and $\av{\sigma_x} = \cos \theta_0$. The substantial deviation of $\av{\sigma_x}$ from $0$ indicates that these parameters lie well outside the regime in which the adiabatic approximation holds. Again, the approximation captures the envelope of the qubit dynamics well, although it does not reproduce the small-amplitude fast oscillations present in the numerical solution. Physically, the large oscillations in $\av{\sigma_z}$ arise because the high-frequency qubit is able to adiabatically follow the tunneling motion of the oscillator~\cite{Ashhab2010}.

\section{Effective potential barrier}

In the previous section we have shown that the Hamiltonian \eqref{e1} exhibits, in the $\epsilon <1$ parameter region, the kind of ``tunneling'' motion normally associated with a double-well system, where the ground state is split into a doublet, with energy difference $\hbar\Delta\omega$, and a state initially localized near the bottom of one of the wells ends up tunneling back and forth at a frequency $\Delta\omega$.  A characteristic feature of this tunneling motion is that the probability to find the system in the region of the potential barrier, that is, right in between the two wells, is always very low, so we could say that the system manages to make it from point A to point B without ever having a significant probability to be found at the intermediate point C.  Formally, there is never a probability peak at the barrier, something which is also ensured by the fact that the wavefunction must be concave at that point, as we shall discuss below.

It is natural to ask if the tunneling-like motion described in the previous section can be understood in terms of some effective potential barrier for the oscillator in our problem.  The very early work of Graham and H\"ohnerbach \cite{Graham1984}, recently revisited by Ashhab and Nori~\cite{Ashhab2010}, does, indeed, show one way in which such an effective potential can be constructed. Here, without necessarily duplicating that work, we wish to present another couple of ways to look at the problem, which is complicated by the fact that the full Hilbert space includes qubit as well as oscillator degrees of freedom.

We begin by briefly summarizing the adiabatic approach of Graham and H\"ohnerbach~\cite{Graham1984}, which provides a useful basis for comparison. Working in position space, the eigenstates of the system may be written in the form $\ket{\psi} = \psi_1(q) \ket{+x} + \psi_2(q) \ket{-x}$, where $q$ is the position coordinate of the oscillator. Inserting this state into the Schrodinger equation yields a pair of coupled differential equations corresponding to the two orthogonal spin components. If the kinetic energy term is negligible (corresponding to the low-frequency/high-inertia limit of the oscillator), the problem reduces to a pair of coupled algebraic equations that may be solved for energy as a function of position. Graham and H\"ohnerbach find two potential energy bands, but as we are primarily interested in low-lying energy levels, we will only look at the lower band, which is given by
\begin{equation}
E_b(q) = \frac{m\omega_0^2}{2}\,q^2 - \sqrt{2\hbar m\omega_0 \lambda^2 q^2 + \frac{\hbar^2\Omega^2}{4}} -\frac{\hbar\omega_0}{2} .
\label{effband}
\end{equation}
This energy band may be thought of as an effective potential for the boson, created by the coupling to the (high-frequency) spin. The point $\lambda^2 = \Omega \omega_0/4$ corresponds to a bifurcation point of the function: for smaller values of $\lambda$ it has a single minimum at $q=0$, but above this critical value the function develops a double-well structure.

The calculation outlined above provides one way of arriving at an effective double-well potential that leads to tunneling-like dynamics of the boson. In the remainder of this section, we present some alternative ways of arriving at this interpretation, together with some analysis of the extent to which the effective potential picture can be justified.

\subsection{Curvature of the probability distribution}

For a single particle in one dimension, it is always possible to obtain the potential $V(q)$ from any energy eigenfunction $\psi_E(q)$, by inverting the Schr\"odinger equation:
\begin{equation}
V(q) = \frac{1}{\psi_E(q)}\left[\frac{\hbar^2}{2 m}\, \frac{d^2\psi_E}{d q^2} \right] + E .
\label{n20}
\end{equation}
In our case, the oscillator generally does not possess a wavefunction of its own, since the states of the oscillator and the qubit are typically entangled. Instead, we may work with the probability distribution $\rho(q)$ for the position of the oscillator, which can always be calculated from the total state vector $\ket\Psi$ as $\rho(q) = \av{\Psi|q}\av{q|\Psi}$.  \emph{If} the oscillator had a separate wavefunction $\psi$ (assumed real for simplicity), then we would have $\rho = \psi^2$, and a little algebra yields
\begin{equation}
\frac{\psi^{\prime\prime}}{\psi} = \frac{\rho^{\prime\prime}}{2\rho} - \left(\frac{\rho^\prime}{2\rho}\right)^2 .
\label{n21}
\end{equation}
We may then work out what the effective potential $V(q)$ would look like by calculating
\begin{equation}
V(q) = \frac{\hbar^2}{2 m}\,\left(\frac{\rho^{\prime\prime}}{2\rho} - \left(\frac{\rho^\prime}{2\rho}\right)^2 \right) + E
\label{n22}
\end{equation}
for several stationary states; if the results obtained for different energies $E$ agree well with each other, this may be taken to support the ``effective potential'' picture.  We note in passing that at the center of the well, by symmetry, $\rho^\prime = 0$, and hence the curvature of $\rho$ determines whether $V(0)$ is greater than $E$ (concave $\rho$) or the opposite.  A positive $\rho^{\prime\prime}(0)$, therefore, is consistent with the picture of a trapped bound state, with energy below the barrier; that is, a conventional tunneling scenario.

It is tempting at this point to try to obtain analytical results by using the approximate eigenstates derived in the previous section.  The form of $\rho_{\pm,0}$ for the ground-state doublet, in particular, is especially simple: up to a normalization factor, one has
\begin{equation}
\rho_{\pm,0}(q) = |\av{q|\alpha_0}|^2 + |\av{q|-\alpha_0}|^2 \pm 2 \cos\theta \,\av{q|\alpha_0}\av{q|-\alpha_0}
\label{n23}
\end{equation}
where $\av{q|\alpha}$ is just the wavefunction of a coherent state with real parameter $\alpha$, given by Eq.~(\ref{e5}).  A little algebra then yields the result 
\begin{equation}
V(0)-E_{\pm,0} = \frac{\hbar\omega_0}{2}\left(\frac{4\alpha_0^2}{1\mp\epsilon}-1\right) \simeq 2\frac{\hbar\lambda^2}{\omega_0} - \frac{\hbar\omega_0}{2} \pm \frac{\hbar\Omega}{2}
\label{n24}
\end{equation}
where the approximation assumes $\epsilon \ll 1$.  One must, however, be wary of trying to extract such sensitive information from what is, after all, only a variational wavefunction; there is, indeed, no guarantee that the curvature of the real $\rho(q)$ is well matched at all by these approximations.  As we shall see below, the result (\ref{n24}) is indeed correct in order of magnitude only.  We may also extract from it an approximate condition to have at least one bound state with energy below the barrier, namely, $2\lambda > \omega_0\sqrt{\omega_0 + \Omega}$; again, a better criterion will be provided below.

\begin{figure}[htbp]
\begin{center}
\includegraphics[]{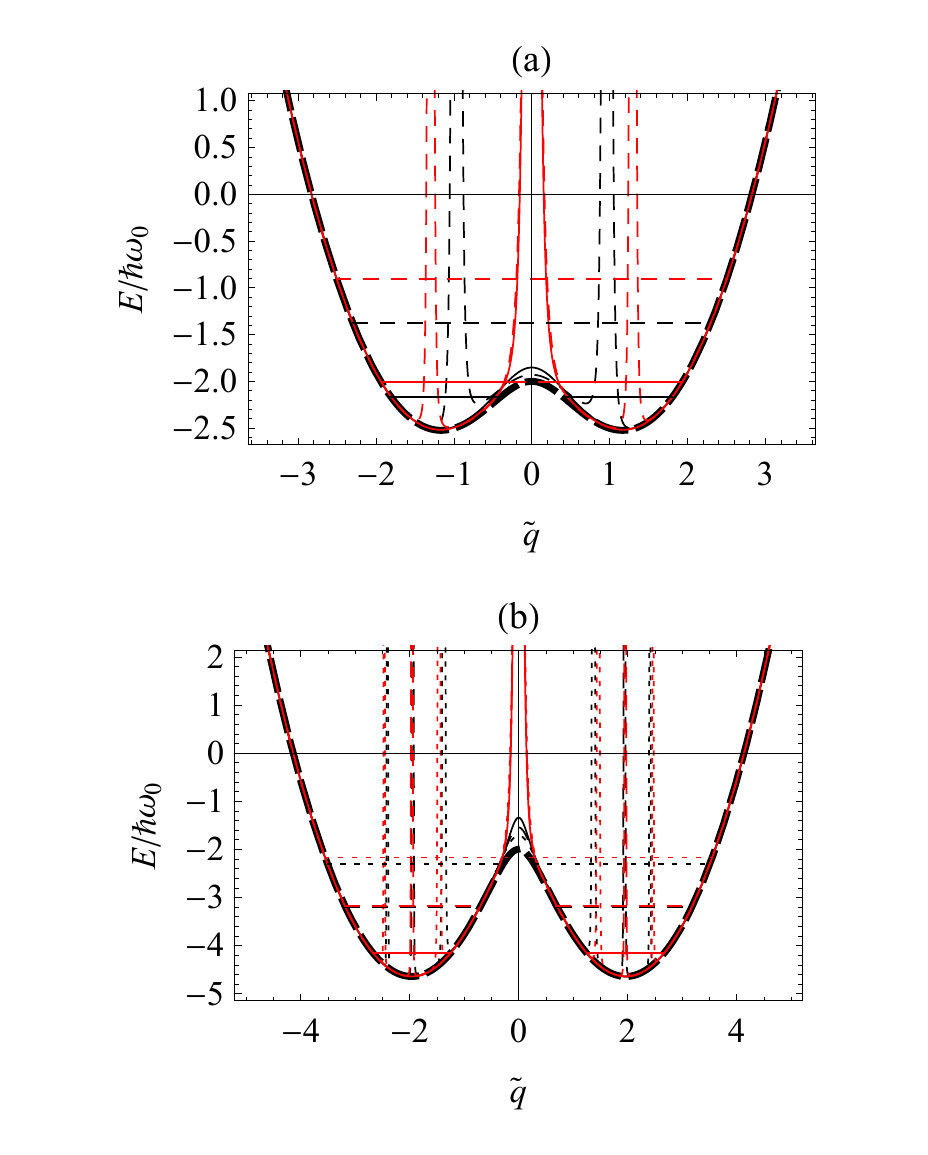}
\caption{(a) Effective potentials, as a function of $\tilde{q} = \sqrt{2 \hbar/m \omega_0} q$, calculated from Eq.~\eqref{n22} for $\Omega=3\omega_0$ and $\lambda = 1.3\omega_0$. Negative (positive) parity results are given by black (red) lines; within each subspace the ground state (first excited state) is indicated by solid (dashed) lines. The numerically calculated energies for the four eigenstates are given by the corresponding horizontal lines. For comparison, $E_b(q)$ (Eq.~\eqref{effband}) is plotted as a thick, dashed black line. (b) Same as (a) but for $\lambda = 2\omega_0$.  Three positive and three negative parity results are shown, given by (solid, dashed, dotted) lines. Note that the energy splitting of the lowest two doublets cannot be resolved on the scale of this figure.}
\label{fig3}
\end{center}
\end{figure}

Instead of the approximate variational eigenstates, Eq.~\eqref{n22} may be evaluated using the numerically calculated eigenstates, which are in principle arbitrarily exact.  For the case $\Omega=3\omega_0$ and $\lambda = 1.3\omega_0$, Fig.~\ref{fig3}(a) shows the result of considering the four lowest eigenstates, two of positive parity (solid lines) and two of negative parity (dashed lines).  The various calculated $V(q)$ agree fairly well, except near the points where $\rho(q)$ almost vanishes (more on this below); moreover, except in these regions, they also agree very  well with the black dashed line, which is the effective potential $E_b(q)$ given by Eq.~\eqref{effband}. Figure~\ref{fig3}(b) shows the case $\Omega=3\omega_0$ and $\lambda = 2\omega_0$.  A total of 6 eigenstates (3 of positive and 3 of negative parity) have been used, and again the agreement between the dashed line representing $E_b(q)$ and the effective potentials calculated via Eq.(\ref{n22}) for all these states is very good except for a few isolated spots (including the center of the barrier, $q=0$).

%

To understand these discrepancies, it should be noted that they they are not actual singularities of Eq.~(\ref{n22}), although they do occur around points where $\rho(q)$ nearly vanishes.  If the oscillator were actually described by the potential energy $E_b(q)$, then these points would be exact nodes of the corresponding eigenfunctions $\psi_E$, and, by Schr\"odinger's equation, $\psi_E$ in the neighborhood of one of these points would have an expansion of the form $\psi_E(q) \simeq a (q-q_0) + b (q-q_0)^3$ with some coefficients $a$ and $b$; then $\rho(q)$ would have the form $\rho(q) \simeq a^2 (q-q_0)^2 + 2ab (q-q_0)^4$, and taking limits in Eq.~\eqref{n22} would yield $V(q_0) \propto 6b/a$.  On the other hand, for the coupled system considered here the actual $\rho(q)$ is not derived from an underlying wavefunction, and there is no reason for it to vanish exactly at $q_0$; rather, it takes the approximate form
\begin{equation}
\rho(q) \simeq c + {a^\prime}^2 (q-q_0)^2 + \ldots
\label{n26}
\end{equation}
with a very small $c$, and substitution in Eq.~\eqref{n22} yields $V(q_0) \propto {a^\prime}^2/c$, where ${a^\prime}^2/c$ is typically very large.
   
Put differently, the plots in Fig.~\ref{fig3} magnify the discrepancies between the exact probability distribution $\rho(q)$ and the solutions to the effective potential $E_b(q)$, but they do that precisely near the ``unimportant'' regions where the probability to find the particle is very small anyway.  This includes, for the lowest energy states, the middle of the potential barrier, $q=0$.  Keeping this in mind, we can assert that the potential $E_b(q)$ of Eq.~(\ref{effband}) does provide a remarkably good approximation, especially when one considers that the oscillator does not even have a true wavefunction of its own, since it is typically in a highly entangled state with the spin.  

A simple calculation shows that the potential $E_b(q)$ does predict, for $\epsilon < 1$, the two minima at the same locations as the variational calculation, $q = \pm(m\omega_0/2\hbar)(\lambda/\omega_0)\sqrt{1-\epsilon^2}$ (compare Eq.~(\ref{e10})). The value of the potential at these minima is
\begin{equation}
\left(E_b\right)_{min} = -\frac{\hbar\lambda^2}{\omega_0}\left(1+\epsilon^2\right) -\frac{\hbar\omega_0}{2}
\label{n27}
\end{equation}
If one assumes that the lowest energy eigenstate will have an energy $\hbar\omega_0/2$ above the bottom of the band, one obtains a good approximation to the ground state energy predicted by the variational method (compare to Eqs. \eqref{e11} and\eqref{e17}).  Analogously, the height of the barrier predicted by $E_b$ is
\begin{equation}
E_b(q=0) = -\frac{\hbar\Omega}{2}-\frac{\hbar\omega_0}{2}
\label{n28}
\end{equation}
Assuming that the lowest levels are spaced by about $\hbar\omega_0$, one can combine these results to predict approximately the number $N$ of tunneling doublets, i.e., pairs of states with energies below the barrier (note that the energy of the $N$-th doublet would be $(N-1/2)\hbar\omega_0$ above the bottom of the band, that is, we start counting states from 1, not from 0):
\begin{equation}
N < \frac{\lambda^2}{\omega_0^2}\left(1+\epsilon^2\right)-\frac{\Omega}{2\omega_0}+\frac{1}{2} .
\label{n29}
\end{equation}
For the cases illustrated in Fig.~3, Eq.~(\ref{n29}) predicts, respectively, $N<1.02$ and $N<3.14$, which agrees with the figures.  An alternative way to predict $N$ is developed in the next subsection.

Finally, we would like to emphasize that the effective potential $E_b(q)$ appears to work well for the higher excited states as well, and not just for the states below the barrier.

\subsection{Effective barrier from the doublet energies}

Typically, in a double-well situation, the lowest-lying energy eigenstates form doublets of closely-spaced energies, where the energy difference gives the rate of tunneling through the barrier; this increases as the overall energy increases and the states move closer to the top of the barrier.  The formula for the energies, Eq.~\eqref{e17}, derived in Section 1 from variational considerations, does indeed exhibit this behavior for the lowest few eigenstates, and one can use this to establish an effective ``barrier height'' as follows.

First, we note that Eq.~\eqref{e17} can be simplified by expanding the normalization factor, since we are typically interested only in situations where both the parameter $\epsilon$ and the overlap factor $e^{-2\alpha_0^2}L_N(4\alpha_0^2)$ are small.  To lowest order in the overlap factor, then, we obtain the simpler result
\begin{equation}
E_{\pm,N} = - \hbar \frac{\Omega}{2} \epsilon + \hbar N\omega_0 - \hbar \frac{\lambda^2}{\omega_0}\left(1 - \epsilon^2\right) \pm \hbar \frac{\Omega}{2}(1-\epsilon^2) e^{-2\alpha_0^2}L_N(4\alpha_0^2) .
\label{n30}
\end{equation}
For the cases illustrated in Fig.~\ref{energies}, the spectra predicted by Eq.~\eqref{n30} are virtually indistinguishable from those predicted by Eq.~\eqref{e17}.

We next observe that the functions $e^{-x}[L_n(x)]^2$ can be used to define probability distributions in the interval $x\in [0,\infty)$, with expectation value $\bar x = 2 n + 1$ and variance $\sigma^2 = 2 n^2 + 2n + 1$.  Observation then shows that $e^{-x/2} L_n(x)$ decays rapidly for $x$ greater than $\bar{x}$ plus about two standard deviations; hence, a condition to have tightly spaced doublets in Eq.~(\ref{n30}) can be expressed as $4\alpha_0^2 \gg 2n+1 + 2 \sqrt{2 n^2 + 2n +1}$, or, again numbering the doublets beginning with $1$ instead of zero, 
\begin{equation}
2 N -1 + 2\sqrt{2 N^2 - 2N + 1} \ll 4\alpha_0^2 .
\label{n31}
\end{equation}
For $N=1,2,3,4$ the left-hand side of (\ref{n31}) has the values $\{3,7.47, 12.2, 17\}$, whereas, for the case depicted in Fig.~3(a), we have $4\alpha_0^2 = 5.38$, and for the case in Fig.~3(b) we have $4\alpha_0^2 = 15.4$.  Hence this equation appears to predict well the number of tunneling doublets in both cases (one in the first instance and three in the second; see Fig.~\ref{fig3}). For large $N$, the left-hand side can be expanded to yield  
\begin{equation}
N \ll 0.83 \frac{\lambda^2}{\omega_0^2}\left(1-\epsilon^2\right) + \frac 1 2 .
\label{n32}
\end{equation}
To leading order, this agrees with all the previous estimates of an effective barrier, whose height at $q=0$ is of the order of $\hbar\lambda^2/\omega_0$, although there clearly are differences between the estimates as well.

From the foregoing considerations, it appears that the predictions of the fully quantized system, both from the approximation developed in Section 2 and from numerical calculations of the full Hamiltonian, are consistent with the semiclassical picture: the interaction with the high-frequency qubit creates an effective potential that takes on a double-well shape for couplings that satisfy $4 \lambda^2/\Omega \omega_0 > 1$. Within this potential, an initial state of the oscillator that is localized in one well tunnels through the barrier and back again. It is worth noting here that the dynamics displayed in Fig.~\ref{oscdynamics} is distinctly different to that of an oscillator coherent state in a single-well potential; see, for comparison, Fig.~1 of Ref.~\onlinecite{Philbin2013}. Our approximation provides a simple but effective means of calculating the tunneling doublet energy splittings and hence the tunneling frequency, as well as predicting how many doublets lie below the energy barrier and thus display tunneling behavior.

\section{Experimental prospects}

The tunneling effect discussed here could potentially be realized in a number of different systems. Perhaps the most exciting experiments from a fundamental point of view would involve a nano- or micro-mechanical resonator as the oscillator component, whose dynamics could then be interpreted as quantum tunneling of a macroscopic object with a direct everyday classical analog. Significant advances over the past few years have shown that this idea is not entirely unrealistic. Mechanical resonators have been cooled very close to the quantum ground state ($
\av{n} \lesssim 1$, where $\av{n} = [\exp(\hbar \omega_0/k_B T) - 1]^{-1}$ is the average number of thermal phonons in a resonator of frequency $\omega_0$ at temperature $T$) by cryogenic techniques~\cite{O'Connell2010} and by sideband cooling via coupling to microwave~\cite{Rocheleau2010,Teufel2011,Safavi-Naeini2012} or optical~\cite{Chan2011,Verhagen2012} fields. Coupling between mechanical resonators and superconducting qubits has been achieved~\cite{LaHaye2009,O'Connell2010,Suh2010,Pirkkalainen2013}, and Rabi oscillations involving the exchange of a single quantized excitation between a qubit and a resonator have been observed~\cite{O'Connell2010}. Further evidence of the quantum nature of a mechanical resonator was provided by a measurement of the distinctively quantum asymmetry in the noise spectrum~\cite{Safavi-Naeini2012}. A few theoretical proposals for creating double-well potentials in which mechanical tunneling could be observed have also been put forward~\cite{Savel'ev2006,Serban2007,Sillanpaa2011,Buchmann2012}.

In trying to set up a tunneling experiment, one faces two conflicting difficulties.  On the one hand, if the tunneling states have energies well below the barrier, so that the probability to find the system in between the two wells is very small, the energy splitting will be exceedingly small, and the tunneling time will become too long for the system to remain undisturbed.  On the other hand, if the states are near the top of the barrier, so that the tunneling time is reasonable, then thermal activation becomes a potential problem that may mask the tunneling signal. 

The time scales for tunneling in the qubit-oscillator system compare well with decoherence times in state-of-the-art mechanical experiments. For a typical superconducting qubit frequency of $\Omega = 10~\text{GHz}$ and a resonator frequency $\omega_0 = \Omega/3 = 3.3~\text{GHz}$, a coupling value of $\lambda/\omega_0 = 1.3$ gives one tunneling doublet near the top of the barrier (see Eq.~\eqref{n31} and Fig.~\ref{fig3}(a)). The time for one transit of the barrier $t_Q = \pi/\Delta \omega$, with $\Delta \omega$ given by Eq.~\eqref{e20}, is then 5.9~ns. This is very close to the resonator energy relaxation time of 6.1~ns measured for a 6~GHz dilatational resonator in the experiments of O'Connell \textit{et al.}~\cite{O'Connell2010}; in the same experiment, the qubit relaxation time was found to be an anomalously short 17~ns. Although high resonator frequency is desirable for shortening the tunneling time, higher frequency comes at the cost of significantly shorter decay times. As another example, take a 100~MHz oscillator with the same 10~GHz qubit. In this case the fractional coupling needed to have one doublet below the barrier is larger, about $\lambda/\omega_0 = 5.1$. The resulting tunneling time is $t_Q=0.22~\mu\text{s}$, which compares favorably with the typical 1~$\mu\text{s}$ decoherence time of superconducting qubits. Resonators with frequencies in the range of 10-100~MHz typically have $Q$ values on the order of $10^5-10^6$~\cite{Rocheleau2010,Teufel2011,Teufel2011a,Palomaki2013}. For these low-frequency oscillators, the important parameter is the rate at which thermal quanta are exchanged with the relatively hot environment, quantified by the thermal decoherence time $\tau_{th} \approx \hbar Q/k_B T_{env}$ where $T_{env}$ is the environment temperature~\cite{Chan2011,Verhagen2012}. While exact values depend on the details of the environment, Palomaki \textit{et al.} estimated $\tau_{th} \approx 90~\mu\text{s}$ in an experiment on a 10.5~MHz resonator~\cite{Palomaki2013}. Therefore the qubit decoherence time is likely to be the limiting factor when a low-frequency resonator is used.  

While a full treatment of thermal effects is beyond the scope of this paper, an estimate of the thermal activation rate $\Gamma_{th}$ can be made using the Arrhenius rate equation~\cite{Sillanpaa2011}: $\Gamma_{th} = \omega_0/(2\pi) \exp(-\Delta V/k_B T)$, where $\Delta V$ is the difference in potential between the bottom of the well and the top of the barrier. The crossover temperature $T_c$ at which the thermal activation rate drops below the quantum tunneling rate can be found by setting $\Gamma_{th} = 1/t_Q$, giving
\begin{equation}
T_c = -\frac{\Delta V}{k_B} \left[ \ln \left( \frac{2 \Delta \omega}{\omega_0} \right) \right]^{-1} ,
\end{equation}
where $\Delta V$ can be estimated from Eqs.~\eqref{n27} and \eqref{n28} and $\Delta \omega$ is given by Eq.~\eqref{e20}. For the first set of parameters considered above, $T_c = 12~\text{mK}$, within the range of modern dilution refrigerators. At lower resonator frequencies the crossover temperature becomes more challenging to achieve; for the second set of parameters above, $T_c = 24~\mu\text{K}$. Thus while lower frequency resonators have the advantage of much higher quality factors and consequently longer thermal decoherence times, high frequency provides a significant advantage in distingushing quantum tunneling from thermal activation over the barrier.


The time scales and temperatures required for observation of quantum tunneling in our scenario are within the reach of current nanomechanics technology. However, one outstanding technical challenge remains, which is achieving the very large qubit-oscillator coupling strength needed to reach the double-well regime. Values for $\lambda/\omega_0$ in current experiments range from about 1\% for the dilatational resonator system~\cite{O'Connell2010} to 5-6\% for flexural resonators~\cite{LaHaye2009,Suh2010,Pirkkalainen2013}. This is about two orders of magnitude smaller than required to create a double-well potential. Suh~\textit{et al.}~\cite{Suh2010} remark that a factor of 10 increase in $\lambda$ should be possible by modifying the geometry. Other types of systems have come closer to achieving the required coupling strength~\cite{Anappara2009,Forn-Diaz2010,Niemczyk2010,Schwartz2011,Scalari2012,Crespi2012}, and a number of proposals for reaching or simulating the ultrastrong and deep strong coupling regimes have recently appeared~\cite{Ballester2012,Romero2012,Grimsmo2013}. This is an area of active research, so further advances are to be expected in the near future.

\section{Conclusions}

We have derived an approximation for the Rabi model in the slow-oscillator regime, $\Omega > \omega_0$, and intermediate coupling strength. As well as giving analytical expressions for the energies and eigenstates of the system in this regime, the approximation allows us to interpret the dynamics of the oscillator and qubit in an intuitive way. An initially localized state of the oscillator displays dynamics similar to that of a particle tunneling in a double-well potential; the high-frequency qubit adiabatically follows the oscillator motion, resulting in slow, large-amplitude oscillations of $\av{\sigma_z}$. This behavior may be interpreted via a semiclassical picture in which the interaction with the qubit creates an effective double-well potential for the oscillator. Within this picture, the fully quantum approximation presented here gives reasonable estimates for the height of the potential barrier, the number of tunneling states trapped below the barrier, and the tunneling frequency of each pair of states. We find that the timescales and temperatures required for realization of qubit-mediated oscillator tunneling are within the reach of cutting-edge micro- and nanomechanics experiments; the only major obstacle is achieving the large coupling strengths required.

\acknowledgments
Funding support from the National Science Foundation and the Leverhulme Trust is acknowledged.

\bibliographystyle{apsrev4-1}

%

\end{document}